# Manipulating the Nanoscale Optical Transmission with a Meta-Transmitarray

Francesco Monticone, Nasim Mohammadi Estakhri, and Andrea Alù[*]

Department of Electrical and Computer Engineering, The University of Texas at Austin

[*]alu@mail.utexas.edu

*By applying the optical nanocircuit concepts to metasurfaces, we propose an effective route to locally control light transmission over a deeply subwavelength scale. This concept realizes the optical equivalent of a transmitarray, whose use is demonstrated for light bending and focusing with unprecedented efficiency over a subwavelength distance, with crucial benefits for nano-optics applications. These findings may lead to large improvements in the manipulation of optical transmission and processing of nanoscale optical signals over conformal and Si-compatible substrates.*

Controlling and molding the transmission of light has been of interest well before the discovery of Maxwell's equations and of the wave properties of light, as lenses have been used to focus and collimate light since the dawn of civilization [1]. Conventional dielectric lenses rely on the accumulation of phase delay during wave propagation, so that the emerging wavefront is tailored in a desired way. One of the major drawbacks limiting the integration of lenses into nanophotonic systems is their significant thickness compared to the wavelength of operation.

This aspect, even more relevant at longer wavelengths, has stimulated the concept of frequency selective surfaces (FSS) [2] at radio-frequencies (RF), which allow controlling reflection and transmission of electromagnetic waves, both in amplitude and phase, over a thin, deeply subwavelength surface. The principle of operation is drastically different from a dielectric lens, since the FSS acts as an ultrathin impedance surface sheet locally modeled with circuit concepts. FSS concepts are commonly used to realize low-profile microwave lenses, and even more flexibility may be achieved combining them in *transmitarrays* [3]-[5], which may be used to reshape the phase front of the transmitted radio-wave by molding the required phase shift as a function of position.

In a different context, the desire to go beyond the refraction properties of conventional materials has led to large interest in engineered metamaterials whose exotic properties may open endless possibilities for applications [6]. Despite the advantages of bulk metamaterials, their fabrication remains a challenge for current technology, especially in the visible range. Therefore, increasing interest has been paid to two-dimensional patterned structures, or metasurfaces, which hold the promise to extend the FSS concepts to optical frequencies, and are especially appealing for their low profile and ease of realization with respect to 3-D bulk structures [7]-[10]. Metasurfaces can provide great flexibility in the design of anomalous reflection and refraction properties. Recent papers [11]-[13] have demonstrated 'generalized Snell's laws' of refraction, obtained by imparting a controlled gradient of phase discontinuity over a single surface using properly designed plasmonic inclusions. There has been fervid discussion over the connection in functionality between the proposed structures and low-frequency transmitarrays and conventional gratings [14]-[15],[16], but the possibility of using a deeply subwavelength surface to control optical transmission at the nanoscale is undoubtedly of interest in a variety of nano-optics applications. Unfortunately, the recently proposed designs have important limitations that hinder their practical application, most importantly the fact that they operate with cross-polarized fields, with inherent limitations on the maximum coupling efficiency that may be achieved. In the following, we introduce the concept of meta-transmitarray, which may be able to translate the much more efficient and powerful solutions available at RF to nanoscale optical devices.

Consider first the case of a single, infinitesimally thin metasurface, transversally inhomogeneous to locally control the transmitted phase, similar to the designs presented in [11]-[13]. These geometries may be successfully modeled in a local sense as an inhomogeneous effective impedance sheet, purely reactive in the lossless case, and sandwiched between two semi-infinite transmission lines with characteristic impedance $Z_0 = 377\ \Omega$ (for normal incidence), as shown in Fig. 1a. The surface impedance is a general function of the transverse position $\mathbf{r}$ on the surface, and this model is accurate as long as the inhomogeneity is not too steep with respect to the wavelength. In order to efficiently manipulate the transmitted beam, it is necessary to achieve minimized reflections, i.e., impedance matching, and a transmission phase controllable over the whole $2\pi$ range. A single interface however can support only tangential electric currents that, due to simple symmetry considerations, radiate equally on both sides, hence drastically limiting the degrees of freedom on tailoring the transmitted fields. On a single surface, in fact, the relation $1 + R = T$ always holds between reflection $R$ and transmission $T$ coefficients, implying that the only way to transmit all the impinging power is to not interfere with it ($R = 0$), leaving no control on the local phase.

In addition, the accessible phase range that can be imparted to the transmitted beam, even with less-than-optimal transmission amplitudes, is still limited by fundamental constraints on the local polarizability of the surface in the scalar scenario. In fact, passivity requires that the imaginary part of the electric polarizability $\alpha_e$ is always positive under an $e^{-i\omega t}$, imposing a severe restriction on the phase of the induced current and radiated fields. To circumvent these limitations, the solutions in [11]-[13] have relied on anisotropic elements, such as V-shaped nanoantennas, that can cover the full $2\pi$ phase span for the cross-polarized terms of the local polarizability tensor. Obviously this approach is not ideal, because the maximum coupling efficiency between the two polarizations is fundamentally limited and, even after maximizing it, symmetry requires radiating half of the coupled energy to the reflection side of the surface. Overall, the efficiency of these proposed structures has been well below 50%. It would be extremely beneficial to design a conformal, low-profile metasurface that can control and manipulate a much larger portion of the impinging power, as commonly achieved at RF. At these lower frequencies, a common approach to break these symmetry limitations consists in exciting also tangential magnetic currents on the surface, which is possible only after introducing a nonnegligible finite thickness to the structure. By being able to control electric and magnetic currents on a surface it is possible to tailor an arbitrary transmission pattern, based on the equivalence theorem [17]. This approach is less appealing in optics, due to the saturation of magnetic effects at visible wavelengths [18].

Here we propose a novel solution to the challenge of manipulating transmitted fields, ideally suited for nano-optics, which uses a stack of planarized metasurfaces similar to the ones that we recently introduced to realize broadband circular polarizers [19], but now introducing a suitably tailored form of transverse inhomogeneity. A stack of three metasurfaces (see Fig. 2a), locally modeled as shunt reactances between transmission-line segments as in Fig. 1b, can provide the three degrees of freedom necessary to independently control the phase of the transmission coefficient while minimizing reflection at the same time. By solving the transmission-line model in Fig. 1b, and assuming a uniform distance $d$ between neighboring metasurfaces, we obtained the required condition on the three reactances to achieve $R = 0$ and $T = e^{i\phi(x)}$, where $\phi(x)$ is the desired phase function along the transverse direction:

$$X_1(x) = X_3(x) = -Z_0 \frac{\mathrm{Sin}(k_d d)\mathrm{Sin}(\phi(x)/2)}{\mathrm{Sin}(k_d d + \phi(x)/2)}, \quad X_2(x) = -Z_0 \frac{\mathrm{Sin}(k_d d)^2}{\mathrm{Sin}(2k_d d) + \mathrm{Sin}(\phi(x))}, \quad (1)$$

where $k_d$ is the propagation constant.

The proposed approach is very general and, if we are able to synthesize the reactance profiles in (1), we may in principle realize an arbitrary phase profile $\phi(x)$ of the transmitted wave at the frequency of interest with 100% efficiency. By choosing different phase patterns, the transmitted beam can be tailored in different ways, e.g., anomalous light bending [11]-[13], focusing and collimation of energy [20]-[21] can be achieved with this technique, among many other possibilities. For instance, in order to bend the incident plane wave in a particular direction, the metasurface should provide a constant phase gradient along the structure, i.e., $\phi(x) = \alpha x$, where $\alpha$ is a function of the desired transmission angle $\theta_t$ as $\alpha = -k_0 \sin(\theta_t)$ for normal incidence. Figs. 1c,d show the required reactance profiles to obtain a transmitted plane wave emerging from the metasurface with an angle of 45 degrees. Inner and outer reactances need to be periodic with a superlattice period $L = |\lambda_0 / \sin(\theta_t)|$.

These impedance profiles may be quite straightforwardly implemented at RF using conventional FSS concepts, providing a competitive alternative to existing transmitarray designs. This is however not the purpose of this Letter, in which we instead focus on the more challenging implementation at optical frequencies, at which conventional circuit elements are not available. In order to realize the required inhomogeneous reactance profile in the optical range, we exploit the nanocircuit paradigm introduced in [21] and experimentally demonstrated in **Error! Reference source not found.**, which allows implementing effective optical impedances by relying on displacement instead of conduction effects. Fig. 2a shows the basic nanocircuit element that we use to implement our impedance profile. It is a composite cubic nanoparticle with a plasmonic portion with negative permittivity $\varepsilon_p$ and a dielectric portion with positive permittivity $\varepsilon_d$. They respectively act as a nanoinductor and a nanocapacitor [21], connected in shunt when the electric field is parallel to the common interface [23]-[25], as shown in the inset of Fig. 2b. If this unit cell is periodically repeated on a 2D grid, as in Fig. 2a, a periodic metasurface with effective impedance (per unit length along the nanorods) is obtained **Error! Reference source not found.**

$$Z_{eff} = Z_{plasm} // Z_{diel} = \frac{i}{\omega h \left[ (l-w)\varepsilon_d + w\varepsilon_p \right]} \quad (2)$$

which is purely imaginary (reactive) in the lossless case. As indicated in Fig. 2a, $h$ and $l$ are the thickness and length of the unit cell, respectively, and $w$ is the width of the plasmonic portion. The beauty and power of this approach is that it allows accessing a broad range of reactance values by locally controlling the filling ratio of the building block. As an example of practical implementation with realistic materials and dimensions, we designed and simulated a metasurface working in the mid-infrared range ($\lambda_0 = 3\,\mu\text{m}$), which may be of strategic interest for defense applications. Nonmetallic plasmonic materials, such as aluminum-doped zinc oxide (AZO) semiconductor, are particularly appealing in this frequency range to realize nanocircuit elements, because of their moderate losses and large tunability [26]. At the frequencies of interest, AZO follows a classic Drude model $\varepsilon_p = \varepsilon_\infty - \frac{\omega_p^2}{\omega(\omega+i\gamma)}$, with $\varepsilon_\infty = 3.3$, $\omega_p = 2\pi\,352.2\,\text{THz}$ and $\gamma = 2\pi\,30.4\,\text{THz}$. If losses are considered, the equivalent circuit should be modified to include a shunt nanoresistance. For the dielectric portion, silicon may be used, whose relative permittivity at 3 $\mu$m is $\varepsilon_d = 11.77$ [27]. The dimensions of the unit cell are indicated in the caption of Fig. 2 and have been chosen to be feasible within current nanotechnology limitations [23], still keeping subwavelength the overall size of the building block. The corresponding equivalent reactance per unit length of the optical metasurface as a function of the width $w$ is shown in Fig. 2b. Around the central resonance, both inductive and capacitive reactances are obtained spanning a broad range of values, and using only two different materials. This is an ideal response to implement our concept of *meta-transmitarray* at optical frequencies.

As discussed above, a symmetric stack of three metasurfaces is necessary to be able to fully manipulate optical transmission and at the same time minimize reflections. We show in Figs. 2c,d the amplitude and phase of $T$ through a periodic stack, infinite in the transverse direction, as a function of the building block width in the external and internal metasurfaces, $w_{out}$ and $w_{in}$, using full-wave simulations. A full phase coverage can be obtained combined with high transmission amplitudes within a significant range of design parameters. By locally modifying the filling ratio of the metasurfaces using these design charts, we may be able to implement the desired transverse phase distribution, still keeping the transmission uniformly very large. For example, in Fig. 2d we identified eight different geometries (white circles), within the region of high transmission amplitude (highlighted by the white line), that cover the entire phase range with steps of 45 degrees. By alternating these geometries along the metasurfaces we can locally tailor the phase in the desired manner, by essentially discretizing in finite steps the desired phase profile.

Figure 3 shows the propagation through the stack for the eight designs identified in Fig. 2d. A linearly-polarized plane wave impinges on each structure at normal incidence, and propagates through it with minimal reflection. The phase of the transmitted wave can be shifted from 0 to 360 degrees, as predicted in the transmission line model of Fig. 1b. It should be stressed that our results are drastically different from the ones presented in [11]-[13], since the meta-transmitarray operates on the same polarization of the incident wave and the total field, not only the cross-polarized component, is modified in the desired manner, with very high efficiency. As a result, almost all the incident energy is transmitted and "processed" by the structure, providing a high throughput which is greatly desirable for practical applications.

Ideally, a wide range of phase patterns can be imprinted to the impinging wave by the meta-transmitarray. For example, in order to realize anomalous refraction, a plane wave should emerge with a desired momentum. The required linear phase function to obtain a transmission angle of 30 degrees for normal excitation is shown in Fig. 4a (top panel). By properly discretizing this profile in 45° steps and arranging the building blocks devised in Fig. 2d in a supercell configuration, we design a meta-transmitarray (Fig. 4a, bottom panel) implementing this desired phase function.

The realized reactance profiles are qualitatively consistent with our transmission-line analysis and when a plane wave impinges on the structure with electric field parallel to the nanorods, the meta-transmitarray provides the additional transverse momentum necessary to bend the wave at the desired angle. The full-wave simulation in Fig. 4b considers the realistic stack and shows that the phase front emerging from the surface is indeed planar and propagates in the desired direction. The coupling efficiency of the anomalous refracted beam is very high, as only a small fraction of energy is reflected or coupled to other diffraction orders. Despite metal losses and the discretization, our simulations predict a coupling efficiency over 75%, and with proper optimization this value may be further increased. The same surface may be also excited at oblique incidence, providing the same additional momentum, leading to a form of "negative" refraction within a certain range of incidence angles. To further demonstrate the flexibility of this approach to control the nanoscale optical transmission, we also implemented a quadratic phase profile, as shown in Fig. 4d, which may be able to focus the impinging radiation at a desired distance, realizing a subwavelength conformal lens, directly integrable into a nanophotonic system or a CMOS camera. The simulated electric field plot and power distribution in Figs. 4e,f clearly demonstrate the lensing effect, with a focal point at distance $2\lambda$ from the meta-transmitarray. Thanks to the complete control of the phase pattern at the nanoscale, focusing at an arbitrary distance may be achieved. The inherently high transmission amplitude and large phase coverage make this planarized lens superior to other proposed solutions for integrated optical lenses, including arrays of plasmonic scatterers [20] or plasmonic nano-slits [21],[28].

Our results show that the meta-transmitarray concept proposed in this Letter may lead to unprecedented flexibility and efficiency in the manipulation of nanoscale optical transmission. The underlying physics of the meta-transmitarray is drastically different from any type of graded-index planar lens. Both conventional and metamaterial-based lenses [29] produce a phase delay due to propagation effects, whereas here the desired phase pattern is "encoded" in a subwavelength building block by locally changing its filling ratio and the associated reactance, at the same time ensuring impedance matching to the host medium. We have demonstrated both deflection and focusing of light at mid-infrared frequencies with a meta-transmitarray made of only two materials, silicon and aluminum-doped zinc oxide. We have chosen these materials for their compatibility with CMOS technology, making the structure appealing in terms of realization and nanophotonic integration. These concepts may also be extended to include a full control of the transmission amplitude, in addition to phase, while still keeping zero reflection, by controlling losses in the design. In general, the incident light may be "processed" by the meta-transmitarray and "molded" in different forms with ideal efficiency, as almost all the impinging energy is coupled into the desired transmitted beam, with ideally zero reflections. Additional features like tunability may be also explored by controlling the material composition with external bias, or using nonlinearities. These designs may also be extended to 2D arrays by modulating the building block geometry in both transverse directions. We are currently exploring several practical scenarios in which our findings may be directly applied, including nanoscale signal processing, holography and solar cells, which all require efficient manipulation of the impinging radiation. We believe that the meta-transmitarray concept, allowing an independent control of transmission amplitude and phase at a deeply subwavelength scale, may pave the way to significant advances in the manipulation and processing of nanoscale optical signals.

**Figures**

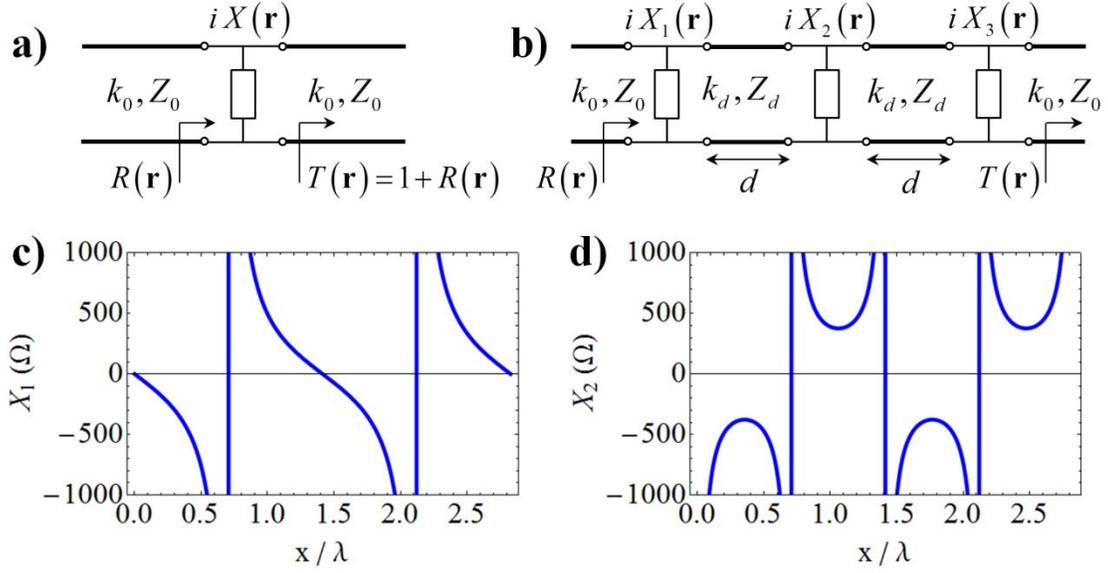

Figure 1 – Transmission line models of (a) a single metasurface and (b) a stack of three metasurfaces separated by dielectric spacers of thickness $d$, under normal incidence. Reactance profiles of (c) the external and (d) internal metasurfaces that implement a linear phase variation along the meta-transmitarray, $\phi(x) = -k_0 \sin(\theta_t) x$. Here we have assumed a transmission angle $\theta_t$ of 45 degrees and an air separation between the metasurfaces of thickness $d = \lambda_0/4$.

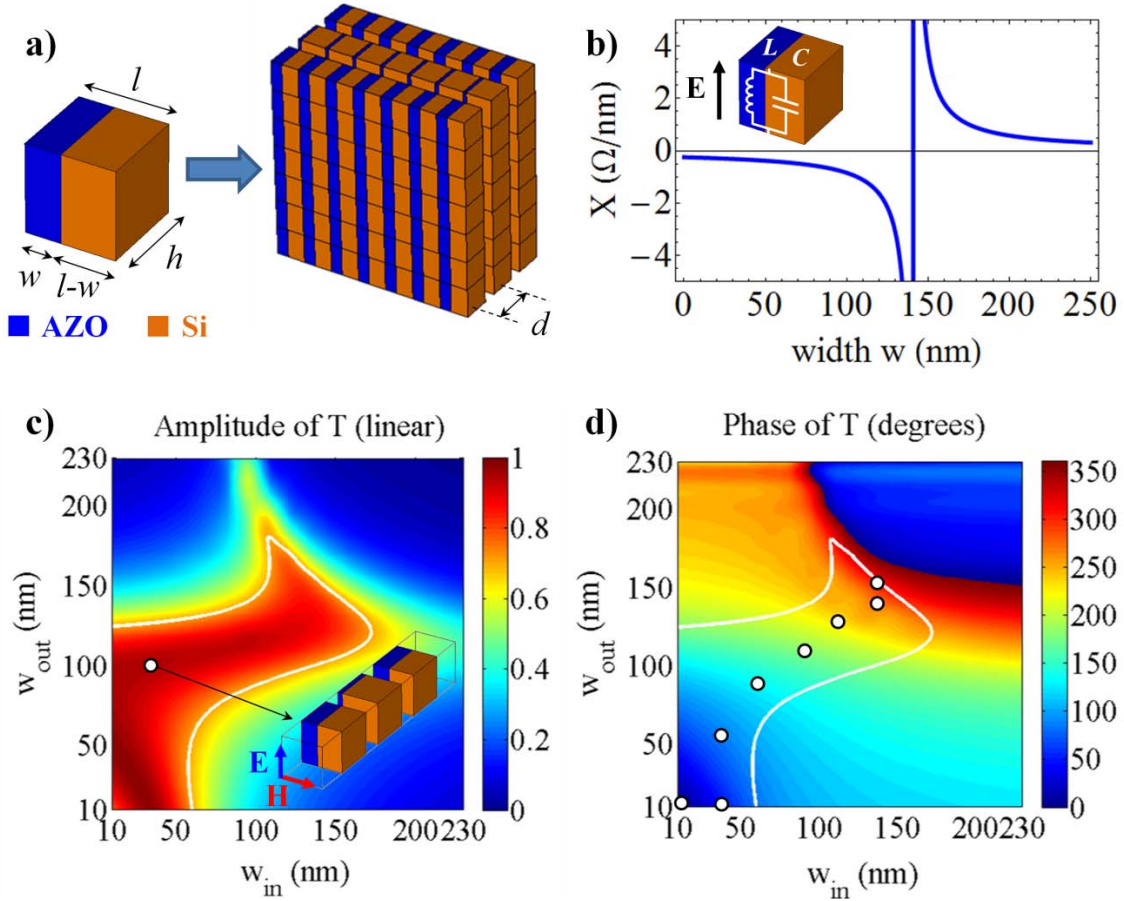

Figure 2 – (a) Basic building block of the metasurface (left) made of plasmonic (aluminum-doped zinc oxide) and dielectric (silicon) materials. Length and thickness of the unit cell are $l = 250$ nm and $h = 250$ nm, respectively. (right) Meta-transmitarray made of three stacked metasurfaces with center-center distance $d = \lambda_0/8 = 375$ nm. (b) Reactance (per unit length) of the nanocircuit element shown in the inset as a function of the width $w$ of the plasmonic nanorod at $\lambda_0 = 3\,\mu$m. (c) Amplitude and (d) phase of the transmission coefficient varying the width of the plasmonic portion in the external and internal metasurfaces. The white contour lines enclose the region with amplitude larger than -3 dB. The inset in (c) shows the geometry corresponding to the white circle. Circles in (d) indicate the eight different configurations used to cover almost 360 degrees of phase shift with steps of 45 degrees.

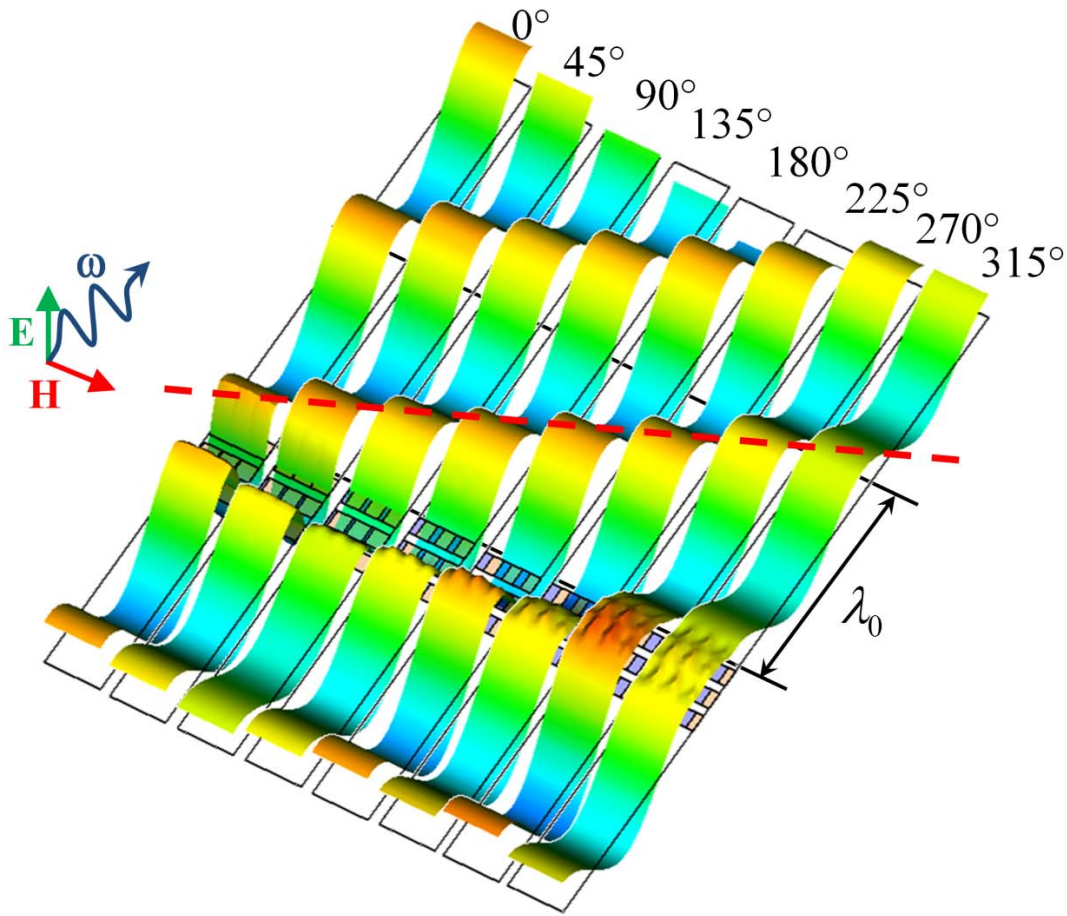

Figure 3 – Propagation through the meta-transmitarray in the eight cases of Fig. 2d. The strips represent the total electric field at the same time instant, clearly showing the different phase shifts obtained for each geometry. Each strip is calculated by an independent full-wave numerical simulation and an impinging linearly polarized plane wave with free-space wavelength $\lambda_0 = 3\,\mu$m is considered in all the cases. The red dashed line follows the peak of the electric field in the eight cases. The position of the peak can be shifted up to a wavelength, which corresponds to a total phase shift of $2\pi$.

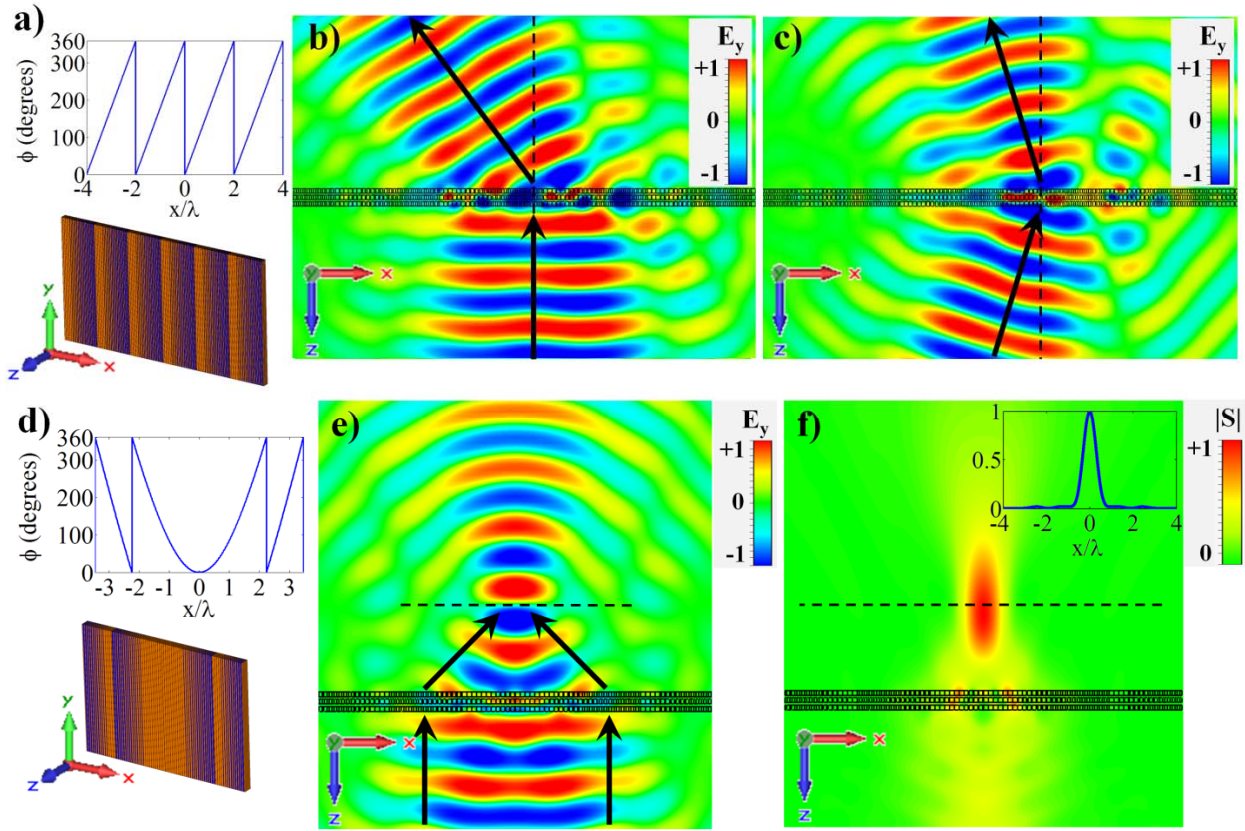

Figure 4 – (a) Linear phase function (top panel) to bend the impinging wave by 30 degrees and (bottom panel) implemented meta-transmitarray. (b) Electric field distribution (snapshot in time) for normal incidence and (c) oblique incidence, for the geometry in (a). (d) Quadratic phase function (top panel) to obtain focusing at $2\lambda$ from the metasurfaces and (bottom panel) implemented meta-transmitarray. (e) Electric field distribution (snapshot in time) and (f) power density distribution (magnitude of the Poynting vector), for the geometry in (d). The dashed horizontal lines indicate the focal plane. The inset in (f) shows the normalized power density along the focal plane. An incident plane wave with Gaussian amplitude profile is assumed in all cases.